\documentclass[
	a4paper, 
	10pt, 
	unnumberedsections, 
	twoside, 
]{LTJournalArticle}

\runninghead{} 

\footertext{\textit{RISC-V Summit Europe, Paris, 12-15th May 2025}} 

\title{Open Challenges for a Production-ready Cloud Environment on top of RISC-V hardware} 

\author{%
	Guillem Senabre\textsuperscript{1}
    \thanks{Corresponding author: \href{guillem.senabre@bsc.es}{\tt guillem.senabre@bsc.es}},  
    Aaron Call\textsuperscript{1}, Ramon Nou\textsuperscript{1}
}

\date{\footnotesize\textsuperscript{\textbf{1}}Barcelona Supercomputing Center} 

\renewcommand{\maketitlehookd}{%
	\begin{abstract}

As part of the Vitamin-V European project, we have built a prototype of a RISC-V cluster managed by OpenStack, with the goal of realizing a functional RISC-V cloud ecosystem. In this poster we explain the hardware and software challenges encountered while porting some elements of OpenStack. We also discuss the current performance gaps that challenge a performance-ready cloud environment over such new ISA, an essential element to fulfill in order to achieve european technological sovereignty. 
	\end{abstract}
}

\begin{document}

\maketitle 
\sloppy

\section{Introduction and motivation}

The RISC-V Instruction Set Architecture (ISA) is at the core of the European Union's technological sovereignty plans. One key initiative in this direction is the European Processor Initiative (EPI)~\cite{epi}, 
which aims to produce processors ready for mass production based on the RISC-V open-source ISA. These processors are intended to be used in various applications, including cloud computing and data centers.

Large-scale supercomputers and datacenters become essential in many scientific applications requiring big-data processing: discovering relevant clinical, social, economic, or environmental indicators at any scale (e.g. 1 PetaByte of genomic datasets~\cite{omics_petabytes}). However, most current computing architectures are proprietary and closed-source technologies such as x86 and ARM, which creates concerns about the reliability of privacy and security. 

The Vitamin-V\cite{vitaminv} project emerged as an effort to build European data centers based on the EPI processor as a reliable alternative to traditional proprietary systems like x86 and ARM. As a part of the Vitamin-V project, we have devoted the last year to developing a functional OpenStack cluster utilizing real hardware instead of emulators. To this end, we are using a hardware platform based on RISC-V development boards. In this poster, we show how we are using Spieed Lichee PI 4A~\cite{licheepi} as a demonstrator in a RISC-V-based cloud environment.

In this poster, we explain some open challenges in RISC-V that need to be addressed to compete with clouds built on top of commodity hardware, as well as the challenges that we have had to overcome in order to build a functional cloud stack on top of RISC-V, and particularly a cloud stack based on the OpenStack framework.

\section{Ecosystem maturity}

\subsection{Hardware}
When transitioning from QEMU-based emulation to physical hardware, the limited availability of commercial RISC-V boards becomes evident. The available options are either relatively expensive or suffer from significant performance limitations. Notably, only one commercially available HPC board~\cite{milkvjupiter} supports vector extensions (RVV1.0). However, its release was delayed by six months, and it sold out within three months despite its 600 € price tag.

\begin{table}[h]
    \centering
    \begin{tabular}{l c c}
        \toprule
        \textbf{Board Name} & \textbf{Cost (\$)} & \textbf{Availability} \\
        \midrule
        HiFive Pro P550 & 400-500  & Limited stock \\
        HiFive Unmatched & 295  & Available \\
        Milk-V Jupiter          & 599  & Sold out \\
        VisionFive 2   & 100  & Available \\
        LicheePi 4A      & 100  & Available \\
        \bottomrule
    \end{tabular}
    \caption{High-Performance RISC-V boards, their prices, and availability.}
    \label{tab:riscv_hpc_boards}
\end{table}

Moreover, the boards have development purposes in mind, not business-ready use cases. So even in those cases where the boards are based on top of powerful chips, there is a significant performance gap. The reasons for this are mostly because RISC-V ISA is yet being properly defined in numerous extensions and many designers have not yet implemented those in real chips. For instance, RVV1.0 extension was ratified just in 2024, and only one implementation is being commercialized. On the other hand, hypervisor extensions are defined but not built on any chips. Other extensions face similar issues.

As the hardware platform we use a RISC-V development board by Sipeed, the Lichee PI 4A owing to their balanced pricing and capabilities. More specifically, the utilized development platform provides a TH1520 RISC-V CPU (4 Threads), 16GB of RAM, 128GB storage. In addition it also provides a dual Gigabit Ethernet a feature particularly interesting when building an OpenStack cluster.

\begin{figure}[h]
    \centering
    \includegraphics[width=0.8\linewidth]{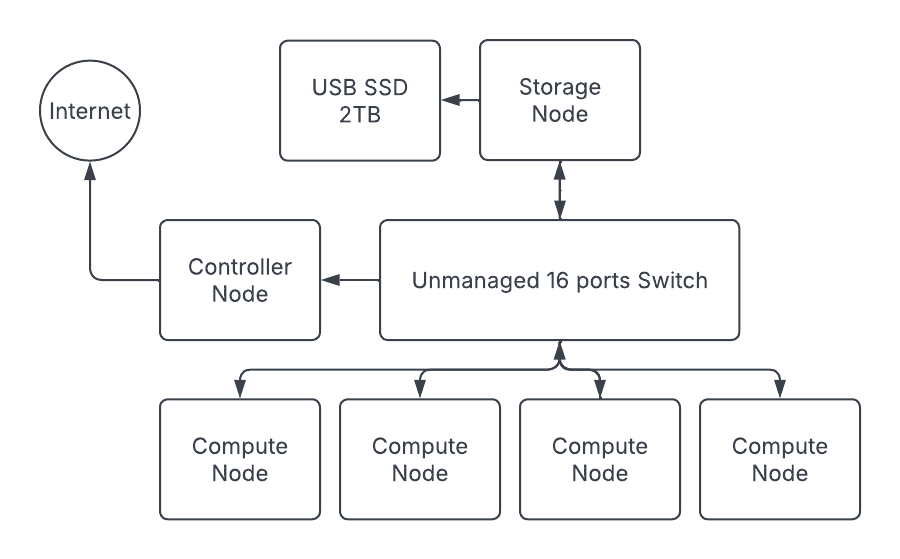}
    \caption{Openstack cluster using LicheePi4a boards.}
    \label{fig:cluster_overview}
\end{figure}

We currently have a controller node, which acts like a bastion, four compute nodes, and a storage node connected to a 2TB USB SSD as shown in figure~\ref{fig:cluster_overview}.

We compared this platform against a cluster composed of 8 nodes based on Intel Xeon Silver 4114 operating at 2.2GHz, with 126GB DDR5 of RAM, interconnected via a 10GBps ethernet network plus a node dedicated to storage with 10TB of capacity. The results of this performance comparison are shown in table \ref{table:performance_comparison}. While it is obvious that a performance gap should exist due to the difference in maturity and performance of both platforms, the gap is way bigger than one could think, in some cases, RISC-V being 10 times slower. These performance issues make it currently unfeasible to enable production-ready cloud environments such as OpenStack, as performance, scalability, and network performance are critical elements. 

\begin{table}[ht]
\centering
\begin{tabular}{|l|c|c|}
\hline
\textbf{Test} & \textbf{RISC-V} & \textbf{x86} \\ \hline
Coremark (iter./sec) & 8500  & 22730 \\ \hline
Mem. Latency 16KB (ns) & 116 & 1497\\ \hline
Disk I/O SeqW. (MB/s) & 204  & 1451 \\ \hline
Disk I/O RandW.(MB/s) & 28.8 & 1316 \\ \hline
Net Throughput (MB/s) & 118 & 1200 \\ \hline
Power Consumption (W) & 11 & 270-350 \\ \hline
\end{tabular}
\caption{Performance Comparison between LicheePi 4A (RISC-V) and Intel Xeon Silver 4112 @ 2GHz (x86) machine.}
\label{table:performance_comparison}
\end{table}

\subsection{Software}

Additionally, the software itself also poses challenges. When enabling OpenStack one realizes that it consists of hundreds of Python libraries and depends on precise package versions. Keeping these packages up-to-date is crucial for stability and compatibility. 

The default Linux distribution for the LicheePi 4A is a Debian-based variant provided by Sipeed, RevyOS, which relies on custom repositories. However, these repositories lacked many essential Debian packages, particularly development tools. Many of those packages are, moreover, not yet ported, so we either had to port them ourselves or wait until the community did. As the project progressed, Debian's official repositories expanded their RISC-V package coverage to approximately 95\%, significantly improving package availability. Consequently, we transitioned entirely to Debian’s repositories to ensure access to the latest software versions.

During OpenStack deployment, we encountered multiple issues related to unported libraries and abandoned dependencies still referenced by other packages. One notable case involved bcrypt and passlib, both of which are used by the OpenStack Identity service (Keystone) for password hashing. 

Moreover, the LicheePi 4A was shipped with kernel 5.10.113, and obtaining an updated image with kernel 6.6.48 required over a month of discussions with the vendor. This slow update cycle highlighted significant maintenance challenges for long-term development and deployment.

Recently, Ubuntu updated its repositories with stable OpenStack packages. Once we discovered this, we started testing a minimal Devstack deployment on a RISC-V Ubuntu image. A patch is still needed to ensure that Nova compute correctly interacts with lower-level libraries like LibVirt or QEMU to determine supported architectures. Although further testing is required, no package mismatches were found, and OpenStack services were installed successfully.

\subsection{Launching Virtual Machines on RISC-V with OpenStack}

Deploying virtual machine (VM) instances on an OpenStack cluster using RISC-V architecture introduced significant challenges, especially when using RevyOS or Ubuntu-based images.

Unlike x86 systems, RISC-V does not feature a standard BIOS. Instead, it typically relies on OpenSBI, which operates in machine mode (M-mode) and transfers control to an externally provided Linux kernel through QEMU (via -kernel) along with a device tree blob (DTB) (via -dtb). This approach complicates the creation of a single, self-contained bootable image, such as the qcow2 format commonly used on x86 systems, which typically embeds a bootloader, kernel, and initramfs.

To address this limitation in an OpenStack environment, we customized the libvirt configuration to inject the kernel and DTB manually at boot time. This is necessary because OpenStack assumes that the guest image is self-bootable and does not natively manage external boot components. A more robust solution involves using EDK2, the UEFI firmware implementation for RISC-V. With EDK2, it becomes possible to create a GPT-partitioned disk image containing an EFI System Partition (ESP) and an EFI executable, such as BOOTRISCV64.EFI or GRUB. This enables the construction of a standard bootable image more in line with OpenStack expectations.

However, UEFI on RISC-V differs from x86 in a critical way: it uses Device Tree for hardware description rather than ACPI. If ACPI is not explicitly disabled, it may conflict with the Device Tree, leading to misbehavior or boot failure. This stems from the fact that ACPI, originally designed for x86, assumes certain platform characteristics that are not present or fully supported on RISC-V.

To ensure proper behavior, ACPI support must be manually disabled in OpenStack. This is typically done by modifying the Nova libvirt driver (e.g., \texttt{nova/virt/libvirt/driver.py}). Additionally, OpenStack's Glance service must be configured with metadata to:

This configuration ensures that the virtual machine boots using EDK2 with the correct hardware description.

Using this approach, we successfully launched multiple VM instances on our OpenStack RISC-V cluster. In the following section, we present a performance comparison between three environments:
\begin{enumerate}
  \item Native RISC-V bare-metal execution
  \item Virtual machines running on QEMU
  \item Instances deployed via OpenStack
\end{enumerate}

\begin{table}[ht]
\centering
\begin{tabular}{|l|c|c|c|}
\hline
\textbf{Metric} & \textbf{Native} & \textbf{OpenStack} & \textbf{QEMU} \\ \hline
real [ms] & 5.81 & 408.00 & 382.00 \\ \hline
user [ms]& 1.78 & 218.70 & 227.80 \\ \hline
sys [ms] & 4.13 & 121.70 & 138.00 \\ \hline
primes [s] & 4.576 & 95.533 & 90.576 \\ \hline
write [s] & 0.042 & 5.955 & 5.762 \\ \hline
read [s] & 0.008 & 0.237 & 0.338 \\ \hline
\end{tabular}
\vspace{1ex}
\caption{Performance metrics obtained from custom C benchmark programs.}
\label{table:cpuinfo_sorted}
\end{table}

\section{Conclusion}

The RISC-V ecosystem is rapidly evolving, as we have observed during the porting of OpenStack, however, it also highlights a lack of robustness and stability in the current state of the platform.
These simple benchmarks, comparing different emulation environments and bare-metal RISC-V, reveal a significant performance gap with the server counterpart architecture x86. In some cases, emulation is up to 100× slower than bare-metal execution. 

In conclusion, the reliance on workarounds and non-standard deployment methods to establish a functioning virtualized environment underscores the current limitations of RISC-V for production-grade cloud infrastructure.
\section*{Acknowledgements}
This work has been partially financed by the European Commission (EU-HORIZON VITAMIN-V GA 101093062), the Spanish Ministry of Science (MICINN) under scholarship BES-2017-081635, the Research State Agency (AEI) and European Regional Development Funds (ERDF/FEDER) under DALEST grant agreement PID2021-126248OB-I00, \newline MCIN/AEI/10.13039/ 501100011033/FEDER and PID GA PID2019-107255GB-C21, and the Generalitat de Catalunya (AGAUR) under grant agreements 2021-SGR-00478 and 2021-SGR-01626.

\bibliography{bibliography}


\end{document}